\def\den{\hbox{den}}
\def\hc{\hbox{h.c.}}

\def\ln{\ell{n}}
\documentstyle[12pt]{article}

 {\parindent 0.5cm \textwidth 15.0cm \textheight
23.5cm \hoffset=-0.8cm \voffset=-3cm
  \let\LARGE=\large
 \let\large=\normalsize
\newcommand{\be}{\begin{equation}}
\newcommand{\ee}{\end{equation}}
\newcommand{\ba}{\begin{array}{c}}
\newcommand{\ea}{\end{array}}

\begin{document}
\begin{titlepage} \vspace{0.2in} \begin{flushright}
MITH-94/9 \\ \end{flushright} \vspace*{1.5cm}
\begin{center} {\LARGE \bf  A Critical Surface of Chiral-invariant System
with Gauge Boson and Fermions\\} \vspace*{0.8cm}
{\bf She-Sheng Xue$^{a)}$}\\ \vspace*{1cm}
INFN - Section of Milan, Via Celoria 16, Milan, Italy\\ \vspace*{1.8cm}
{\bf   Abstract  \\ } \end{center} \indent

In the chirally-invariant context of the $U_{em}(1)$ gauge interaction and
four-fermion interactions for ordinary and mirror fermions, the Schwinger-Dyson
equation for the fermion self-energy function is studied on a lattice. We find
that a sensible infrared limit can be defined on a critical surface, which is
consistent with the critical line found in the continuum theory.
\vfill \begin{flushleft} 15th February, 1994   \vspace*{3cm} \\
\noindent{\rule[-.3cm]{5cm}{.02cm}} \\
\vspace*{0.2cm} \hspace*{0.5cm} ${}^{a)}$
E-mail address: xue@milano.infn.it\end{flushleft} \end{titlepage}

\section{Introduction}\indent
Dynamical symmetry breaking plays an important role in
understanding success of the Standard Model. The structure of dynamical
symmetry breaking has been studied intensively for a wide variety of
theoretical models. Due to the lack of clear ideas as to the physical origin
of dynamical symmetry breaking in the Standard Model and the essential
difficulties in attempting to study such non-perturbative phenomena, progress
is still quite limited. One theoretical approach to the study of this
phenomenon is to find the solutions of the Schwinger-Dyson equation (SDE) for
the
fermion self-energy function $\Sigma(p)$. For a homogeneous SDE
without an explicit ultraviolet cutoff ($\Lambda$), it was shown \cite{John}
that the mass operator ($\bar\psi\psi$) has an anomalous dimension
$d_{\bar\psi\psi}=2+\sqrt{1-3\alpha/\pi}$ for a finite gauge coupling and thus
the high momentum decrease of solutions $\Sigma(p)\rightarrow
(p^2)^{-3+d_{\bar\psi\psi}}$ as $p^2\rightarrow\infty$. Further work
\cite{maskawa} in a version of the theory with an explicit ultraviolet cutoff
shows: (i) there are (no) spontaneous symmetry breaking solutions to the
homogeneous SDE for gauge coupling $\alpha > {\pi\over 3}$ ($\alpha <
{\pi\over 3}$); (ii) for an inhomogeneous SDE, all solutions for weak
coupling require that the inhomogeneous term, i.e. an explicit bare fermion
mass
$m_0$, goes as $m_0\sim \Lambda^{-(1-\sqrt{1-3\alpha/\pi})}$ in the
continuum limit ${\Lambda\over m}\gg 1$, where $m$ is an infrared
mass scale for the theory. However the mass operator
$m_\circ\bar\psi\psi$ remains finite at this limit and generates explicit
breaking of chiral symmetry.

The solutions to the SDE were analyzed in detail for both weak and strong
couplings \cite{kugo}. The critical point $\alpha_c=\pi/3$ is viewed
\cite{mira} as an ultraviolet fixed point of the theory in order to give an
sensible infrared limit for the theory. However, the renormalization properties
of the theory are called into question for the required running of the gauge
coupling with the cutoff. The most important progress made by Bardeen, Leung
and Love (BLL)\cite{bardeen1} is to consider a system of gauge interaction and
four-fermion interaction, which is induced by integrating out the
high-frequence contribution of the theory. The theory was shown to exhibit a
line of critical points separating the chirally broken phase from the symmetric
one. The renormalization properties of the theory, the relevance of the
four-fermion operators and the spontaneous breaking of chiral symmetry with
accompanying pseudo-Goldstone bosons were intensively studied. Recently, in
Ref.\cite{xue1}, we consider this problem on a lattice by adding the Wilson
fermion and bare mass terms, which however are explicitly chirally-variant. In
this paper,
using the gauge-invariant and chirally-invariant lattice regularization, we
study the SDE for the fermion self-energy function in the system containing
gauge interaction and four-fermion interactions for both ordinary fermions and
mirror fermions. As an initial step, we show a critical surface, where a
sensible infrared limit of the theory can probably be realized.

\section{Action}

The fermion ``doubling'' phenomenon is a well-known
problem arising when fermion fields are defined on a lattice \cite{nogo}.
Wilson \cite{wilson}, however, introduced
an extra dimension-5 operator (the Wilson term) into the naive lattice
pure gauge $S_g(U)$ and the Dirac action $S_d(\bar\psi,\psi,U)$ for
ordinary and mirror fermions,
\begin{equation}
S=S_g(U)+S_d(\bar\psi,\psi,U)
+{r\over a}\sum_{x,\mu}\bar\psi(x)\partial^2_\mu\psi(x)),
\label{wilson}
\end{equation}
where $a={\pi\over\Lambda}$ is the lattice spacing and
the lattice ``laplacian'' $\partial_\mu^2$ is defined as
\begin{equation}
\partial_\mu^2\psi(x) =U_{\mu}(x)\psi(x+a_\mu)
+U^\dagger_\mu\psi(x-a_\mu)-2\psi(x).
\label{laplacian}
\end{equation}
This extra term with a finite Wilson parameter $r$ ($0 < r \leq 1$ for
reflection positivity) becomes an irrelevant operator for ordinary fermions
with $ka\sim 0$ in the infrared limit while it is a relevant operator for
mirror fermions with $ka\sim \pi$ in the high-energy regime, in fact it
generates an effective mass $M\sim {r\over a}$ for mirror fermions. However,
the Wilson term explicitly breaks chiral symmetry.

{}From a dynamical viewpoint,
we have not understood completely the operator content
of the theory at short distances, where there probably exist local and
non-local high-dimension operators for ordinary and mirror fermions and gauge
bosons, owing to the experimental observation of a rich mass spectrum of
fundamental particles in the low-energy region. At long distances, these
high-dimension operators, however, should be relevant and irrelevant for
ordinary fermions and mirror fermions respectively in such a way that mirror
fermions decouple from the low-energy spectrum and ordinary fermions remain and
couple properly with gauge bosons.
We thus consider relevant and irrelevant high-dimension operators for ordinary
and
mirror fermions respectively. In general, there
are several possibilities of chirally invariant four fermion interactions on a
lattice space-time, e.g.,
\begin{eqnarray}
&&\beta_1\bar\psi(x)\psi(x)\bar\psi(x)\psi(x)+
\beta_1\sum_\mu\bar\psi(x\pm\mu)\psi(x)\bar\psi(x)\psi(x)\nonumber\\
&&+\beta_3\sum_{\mu\nu}\bar\psi(x\pm\mu)\psi(x\pm\nu)\bar\psi(x)\psi(x)
+\cdot\cdot\cdot,
\label{four}
\end{eqnarray}
which represent complicated interactions between ordinary fermions and mirror
fermions. The origin of these interactions (which might stem from the quantum
gravity \cite{xue,xue3,planck} or a high-frequency contribution of the theory
\cite{bardeen1,bardeen2} and other unknown physical dynamics \cite{wei,ep})
will not be a focus of this paper. We have tuned the
four-fermion couplings in eq.(\ref{four}) such that
the interactions between ordinary
fermions and mirror fermions are of Nambu-Jona Lasinio type (NJL) \cite{nambu}.
Thus we consider the following chirally invariant lagrangian
with lattice regularization\cite{xue2}
\begin{eqnarray}
S&=&S_g(U)+S_d(\bar\psi,\psi,U)+S_r+S_{ir}\nonumber\\
S_r&=&-G_1\sum_x\big(\bar\psi_L(x)\psi_R(x)\bar\psi_R(x)\psi_L(x)\big)
\label{action}\\
S_{ir}&=&-{G_2\over 2}\sum_{x\mu}\big(\bar\psi_L(x)\partial_\mu^2\psi_R(x)
\bar\psi_R(x)\partial_\mu^2\psi_L(x)\big),
\nonumber
\end{eqnarray}
where $U\in U_{em}(1)$ is chosen. In eq.~(\ref{action}), the third
term $S_r$ represents the (NJL) interaction \cite{nambu} of ordinary fermions
and mirror fermions. The fourth term $S_{ir}$ is the NJL interaction of mirror
fermions only, where the gauge link connects neighbouring right-handed and
left-handed fermions to have the $U_{em}(1)$ local gauge symmetry
(\ref{laplacian}).
$G_{1,2}$ are two, as yet unspecified, Fermi-type $O(a^2)$ coupling constants.
Note that (i) in the naive
continuum limit, probed by momenta $pa\ll 1$ ($S_{ir}\simeq 0$),
eq.~(\ref{action}) is just a gauged NJL model for ordinary fermions
\cite{bardeen1}; (ii) in the naive ``lattice limit'', probed by momenta
$pa\simeq 1$, $S_{ir}$ is significantly non-vanishing and eq.(\ref{action}) can
be considered as gauged NJL model for mirror fermions \cite{xue4}; (iii)
obviously, $S_{r}
(S_{ir})$ is a relevant (irrelevant) operator for ordinary fermions and both
$S_{r}, S_{ir}$ are relevant operators for mirror fermions.

In eq.(\ref{action}), chiral symmetry is perfectly conserved at short distance,
the point is, however, whether we can separate ordinary fermions from mirror
fermions
and have a sensible infrared limit ${\Lambda\over m}\gg 1$ of the theory.

\section{Gap equations}

We are thus led to study the SDE of
the fermion self-energy function $\Sigma(p)$. The Landau mean-field
method and the quenched and planar approximations (
large-$N_f$ approach $G_{1,2}N_f$ fixed, $N_f \gg 1$ is the number
of flavour in eq.(\ref{action})) will be adopted.
Given the (\ref{action}) with the quadrilinear $S_r$ and $S_{ir}$ terms,
we have, as illustrated in Fig.1,
\begin{eqnarray}
\Sigma(p)&=& 2g_1\int_q{\Sigma(q)+{r\over a}w(q)\over s^2(q)+M(q)^2}
+{\lambda\over a}\int_q
{1\over 4s^2({p-q\over 2})}\left(\delta_{\mu\nu}-\xi{s_\mu({p-q\over 2}))
s_\nu({p-q\over 2})\over
s^2({p-q\over 2})}\right)\nonumber\\
&&\cdot\Big(V^{(2)}_{\mu\nu}(p,p)-V^{(1)}_\mu(p,q){1\over \gamma_\rho s_\rho(q)
+M(q)}V_\nu^{(1)}(p,q)\Big),
\label{sd}
\end{eqnarray}
where $p (q)$ are dimensionless external (internal) momenta;
$s_\mu(l)=\sin(l_\mu)$ and $s^2(l)=\sum_\mu\sin(l_\mu)$; $\lambda=e^2
$ for $U_{em}(1)$ gauge group and $g_1a^2=G_1N_f$; the
Wilson term $w(q)=\sum_\mu(1-\cos q_\mu)$ and $M(q)=a\Sigma(q)+rw(q)$;
$\int_q=\int^\pi_\pi {d^4q\over (2\pi)^4}$. The vertices \cite{smit} are
($k_\mu={(p_\mu+q_\mu)\over 2}$)
\begin{equation}
V_\mu^{(1)}(p,q)=\left(\gamma_\mu\cos k_\mu+
r\sin k_\mu\right);\hskip0.2cm
V_{\mu\nu}^{(2)}(p,q)=a\left(-\gamma_\mu\sin k_\mu+
r\cos k_\mu\right)\delta_{\mu\nu}.
\label{vertex}
\end{equation}
The Wilson parameter $r$ in above equation turns out to be a symmetry-breaking
v.e.v.~($r=\bar r a$)
\begin{equation}
\bar r={G_2\over 4}\!\sum_{\mu,x}\left({1\over 4}\right )
\left\langle\bar\psi_L(x)\partial_\mu^2\psi_R(x)+\hc\right\rangle
,\label{r}
\end{equation}
which obeys gap-equation generated by the NJL self-interaction $S_{ir}$ of
only mirror fermions (\ref{action}),
\begin{equation}
r = {g_2\over 2} \int_q w(q) { \Sigma(q)a+rw(q) \over s^2(q)
+(\Sigma(q)a+rw(q))^2},
\label{gap2}
\end{equation}
where $g_2 a^2 = N_f G_2$ and we eliminate the interaction between gauge
boson and mirror fermions. This gap equation (\ref{gap2}) is the result of
leading order in the large-$N_f$ expansion.

One clearly finds that both ordinary fermions and mirror fermions contribute to
the gap equation (\ref{sd}) for the fermion self-energy function $\Sigma(p)$.
What we
should do is to find a consistent solution to the SDE (\ref{sd},\ref{gap2})
where ordinary fermions can separated from mirror fermions and a sensible
continuum limit can be defined.

\section{Ordinary and mirror fermions}

One of the main novelties of (\ref{sd}) is the non trivial interplay between
the continuum region, i.e., for ordinary fermions with momenta ($q\ll 1$),
and the truly discrete region for mirror fermions $q\simeq 1$. For the small
external momenta $p=p^\prime a\ll 1$, we rewrite
\begin{equation}
\Sigma(p)=\Sigma_c(p^\prime)+{\Delta\over a},\label{s}
\end{equation}
where $\Sigma_c(p^\prime)\,(\Sigma_c(p^\prime)a\ll 1)$ is the self-energy
function of
ordinary fermions in the
continuum theory (region) and $({\Delta\over a})$ is the divergent
contribution stemming from mirror fermions (\ref{sd}).
In order to study such interplay between ordinary-fermion and mirror-fermion
contributions, it is important to
introduce a ``dividing scale'' $\epsilon$, such that $p^\prime a\ll\epsilon\ll
\pi$.
Separating the integration region in (\ref{sd}) into two regions the
``continuum region'' $(0,\epsilon)^4$ and the ``lattice region''
$(\epsilon,\pi)^4$, we may separate our integral equations into the
``continuum part'' and the ``lattice part''
\begin{eqnarray}
\Sigma(p)&=&2G_1\int_{\epsilon\Lambda}{d^4q'\over (2\pi)^4}
{\Sigma_c(q')\over (q')^2+\Sigma_c(q')^2}
+2g_1\beta_1(r,\epsilon)+{r\over a}\beta_2(r,\epsilon)\nonumber\\
&&+{\alpha\over \alpha_c}\int_{\epsilon\Lambda}
{d^4q'\over 4\pi^2}{1\over
(p'-q')^2}{\Sigma_c(q')\over (q')^2+\Sigma_c(q')^2}
+{\alpha\over \alpha_c}\delta_1(r,\epsilon)+{r\over a}\delta_2(r,\epsilon),
\label{separation}
\end{eqnarray}
where $p',q'$ are dimensionful momenta, $\alpha={\lambda\over 4\pi}$ and $
\alpha_c={\pi\over 3}$. The ``continuum part'' of eq.(\ref{separation}), where
the Landau gauge $\xi=1$
is chosen, is same as that derived from the continuum theory with an
intermediate cutoff $\epsilon\Lambda$. As
for the contributions to the integral equation from
the discrete ``lattice region''
$\beta_i(r,\epsilon),\delta_i(r,\epsilon) (i=1,2)$, we obtain:
\begin{eqnarray}
\beta_1(r,\epsilon)&=& \int_{q\in (\epsilon,\pi)^4}
{\Sigma_c(q)\over s^2(q)+M(q)^2}\label{b1}\\
\beta_2(r,\epsilon)&=& 2g_1\int_{q\in (\epsilon,\pi)^4}
{w(q)\over s^2(q)+M(q)^2}\label{b2}\\
\delta_1(r,\epsilon)&\simeq&
-\int_{(\epsilon,\pi)^4}{d^4q\over 4\pi^2}{\Sigma_c(q)\over
4s^2({q\over 2})}
\left[{-c^2({q\over 2})+r^2s^2({q\over 2})\over
s^2(q)+M^2(q)}\right]\label{b3}\\
\delta_2(r,\epsilon)&\simeq& {\alpha\over\alpha_c}\int_{q\in (\epsilon,\pi)^4}
{d^4q\over 12\pi^2}\!{1\over (4s^2({q\over 2}))}\left[{1\over 2}-
{w(q)(-c^2({q\over 2})+r^2s^2({q\over 2}))+s^2
(q)\over s^2(q)+M^2(q)}\right]\label{b4}
\end{eqnarray}
where $c^2(l)=\sum_\mu\cos^2(l_\mu)$.
The dependence on the
external
momentum $p\in (0,\epsilon)^4$ is omitted in $\delta_i(r,\epsilon),
$ because $p\ll q$ in the ``lattice region'' $q\in (\epsilon,\pi)^4$.

Note that (i) $\delta_i(r,\epsilon)$ do not depend on the gauge parameter $\xi$
for there is a perfect cancellation between the ``contact'' and the ``rainbow''
diagrams (Fig.1), which is guaranteed by Ward's identities; (ii) the limits
$\lim_{\epsilon\rightarrow 0}\delta_2(r,\epsilon)$ and
$\lim_{\epsilon\rightarrow 0}\beta_2(r,\epsilon)$ can be taken since the
functions $\delta_2(r,\epsilon)$ and $\beta_2(r,\epsilon)$ are regular in the
limit $\epsilon\rightarrow 0$, while this is not case for the functions
$\delta_1(r,\epsilon)$ and $\beta_1(r,\epsilon)$; (iii) within the ``lattice
region'', the Wilson parameter $r$ must not vanish and all functions
$\beta_i(r,\epsilon)$ and $\delta_i(r,\epsilon)$ remain non-vanishing. The
functions $\delta_2(r,\epsilon)\,(\delta_1(r,\epsilon))$ and
$\beta_2(r,\epsilon)\, (\beta_2(r,\epsilon))$ can be regarded as
the divergent (finite) mirror-fermion contributions to gap equation
(\ref{sd}) through the four-fermion interaction $S_r$ and the gauge interaction
respectively.

In order to find the sensible ``continuum limit'', where the spectrum of mirror
fermions is far separated ($r\sim O(1)$) from that of the observed ordinary
fermions ($\Sigma_c(p^\prime)a\ll 1$), we should be allowed to tune {\bf\it
only} one parameter, which is related to mass renormalization counterterm. Thus
we tune $\Delta$, in such a way that the ``${1\over a}$'' terms in both sides
of (\ref{sd}) agree. It is self-consistent that this tuning is performed
simultaneously in the numerators and denominators of the RHS of
eqs.~(\ref{sd},\ref{gap2}). These observations mean that we can rewrite the gap
equations (\ref{sd},\ref{gap2}) as the following three self-consistent
gap-equations:
\begin{eqnarray}
\Sigma_c(p^\prime)&=&2G_1\int_{\epsilon\Lambda}{d^4q'\over (2\pi)^4}
{\Sigma_c(q')\over (q')^2+\Sigma_c(q')^2}
+2g_1\beta_1(r,\epsilon)\nonumber\\
&&+{\alpha\over \alpha_c}\int_{\epsilon\Lambda}
{d^4q'\over 4\pi^2}{1\over
(p'-q')^2}{\Sigma_c(q')\over (q')^2+\Sigma_c(q')^2}
+{\alpha\over \alpha_c}\delta_1(r,\epsilon)
\label{continuum}\\
{\Delta\over a}&=&{r\over a}\beta_2(r,0)+{r\over a}\delta_2(r,0).
\label{fine}\\
r &=& {g_2\over 2} \int_q w(q) {\Sigma_c(q)a + r w(q) \over \den(q)},
\label{gap5}
\end{eqnarray}
where $\den(q) = \sin^2q_\mu+(\Sigma_c(q)a + rw(q))^2$ and also the denominator
$s^2(q)+M^2(q)$ of eqs.(\ref{b1},\ref{b2},\ref{b3},\ref{b4}) is substituted by
$``\den(q)$'' and thus is free from $({1\over a})$ divergence. The
eq.~(\ref{continuum}) is the ``continuum'' counterpart of the SDE (\ref{sd}) on
the lattice. For the ``lattice'' equation (\ref{gap5}), one clearly finds that
the solution $r > 0$
stems from the contribution of mirror fermions, owing to the factor $w(l)$,
which
does not vanish in the ``lattice'' region.

The consistency of this fine-tuning (\ref{fine}), which can also be called the
``chiral limit'', can probably be guaranteed \cite{rome} by Ward-type
identities due to chiral symmetry of the action (\ref{action}) at short
distances and we shall not discuss it in this paper. The term ${\Delta\over a}$
plays a r\^ole akin to that of mass counterterms \cite{rome}. Its actual value
depends not only on the RHS of eq.~(\ref{fine}), which is the mirror-fermion
contribution at one-loop level, but also all possible $({1\over a})$
contribution from mirror fermions in the SDE (\ref{sd}). Looked at from this
point of view, the
self-consistency of this one-parameter tuning should not be more surprising
than the well-known and checked self-consistency of mass-renormalization in
continuum Quantum Field Theory.

\section{The critical surface}

Let us now address the important question of the $\epsilon$-independence of our
results. The introduction of the ``dividing scale'' in (\ref{sd}) is,
apart from
the requirement $p^\prime a\ll\epsilon\ll\pi$, rather arbitrary, thus no
dependence
on $\epsilon$ should appear in our final results.
In order for such independence to occur, as it must, it
is clear that the $\epsilon$-dependent terms (e.g., $\ln\epsilon$) from the
continuum integral in
(\ref{continuum}), must be compensated by analogous
terms arising in the calculation of $\delta_1(r,\epsilon)$ and
$\beta_1(r,\epsilon)$. Owing to integral momenta $q\in (\epsilon,\pi)^4$ in
eqs.(\ref{b2}),(\ref{b4}), we can make the reasonable
approximation $\Sigma_c(q)\simeq\Sigma_c(\Lambda)$ in the numerators of
$\delta_1(r,\epsilon)$ and
$\beta_1(r,\epsilon)$. Thus the $\epsilon$-independent terms
$\Sigma_c(\Lambda)\delta_0(r) (\Sigma_c(\Lambda)\beta_0(r))$ contained in
$\delta_1(r,\epsilon) (\beta_1(r,\epsilon))$ can be found by
numerical calculation. The numerical functions $\delta_0(r)$ and $\beta_0(r)$
are reported in Fig.2 and 3. Thus segregating
the $\epsilon$-dependent terms in (\ref{continuum}), we may write
\begin{eqnarray}
\Sigma_c(p^\prime)&=&2G_1\int_{\Lambda}{d^4q'\over (2\pi)^4}
{\Sigma_c(q')\over (q')^2+\Sigma_c(q')^2}
+2g_1\Sigma_c(\Lambda)\beta_0(r)\nonumber\\
&&+{\alpha\over \alpha_c}\int_{\Lambda}
{d^4q'\over 4\pi^2}{1\over
(p'-q')^2}{\Sigma_c(q')\over (q')^2+\Sigma_c(q')^2}
+{\alpha\over \alpha_c}\Sigma_c(\Lambda)\beta_0(r),
\label{fingap}
\end{eqnarray}
which is analogous to the ``chiral limit'' SDE of the continuum theory
with two additional boundary terms $2g_1\Sigma_c(\Lambda)\beta_0(r)$ and
${\alpha\over \alpha_c}\Sigma_c(\Lambda)\delta_0(r)$.

It is straightforward to adopt the analysis of
Bardeen, Leung and
Love \cite{bardeen1,bardeen2,kondo}. After performing the angular integration
and changing
variables to $x=(p')^2$, eq.(\ref{fingap}) becomes a boundary-value problem,
\begin{eqnarray}
{d\over dx}(x^2\Sigma_c '(x))+{\alpha\over 4\alpha_c}{x\over x+\Sigma_c^2(x)}
\Sigma_c(x)&=&0\nonumber\\
(1+\tilde g_1)\Lambda^2\Sigma_c '(\Lambda)+\Sigma_c(\Lambda)(1-2\tilde g_1
{\alpha\over\alpha_c}\beta_0(r)
-{\alpha\over \alpha_c}\delta_0(r))&=&0,
\label{diff}
\end{eqnarray}
where $\tilde g_1=G_1a^{-2}N_f{\alpha_c\over\alpha}$. The solution to this
boundary
value problem is well established. For weak coupling, we
have the gap equation,
\begin{eqnarray}
\tanh\theta &=& {(\tilde g_1+1)\sqrt{1-{\alpha\over\alpha_c}}\over
(\tilde g_1-1)+4\tilde
g_1{\alpha\over\alpha_c}\beta_0(r)+2{\alpha\over\alpha_c}
\delta_0(r)}\\
\theta &=& \sqrt{1-{\alpha\over\alpha_c}}\ln\big({\Lambda \over m}\big),
\label{gap}
\end{eqnarray}
where $m=\Sigma_c(0)$ is the infrared scale. In the infrared limit,
${\Lambda\over m}\gg 1$
and $\theta\gg 1$, the critical surface relates $\tilde g_1$, $r$
and $\alpha$ as
\begin{equation}
\tilde g_1 = {(1+\sqrt{1-{\alpha\over\alpha_c}})-
2{\alpha\over\alpha_c}\delta_0(r)
\over (1-\sqrt{1-{\alpha\over\alpha_c}})+4{\alpha\over\alpha_c}\beta_0(r)}.
\label{critical}
\end{equation}
The Wilson parameter $r(g_2)$ is a function of the four-fermi coupling $g_2$.
This function can be approximately found from eq.(\ref{gap5}) for
$a\Sigma_c(q)\simeq 0$ and $\den(q)\simeq \sin^2q_\mu+(rw(q))^2$.
Numerical calculation shows, for $g_2 > 0.2$, there exist
a non-trivial solution $r > 0$, which is reported in Fig.~4.

Based on eq.(\ref{critical}) and functions $\delta_0(r)$, $\beta_0(r)$ and
$r(g_2)$, we obtain the critical surface, which is shown in Fig.~5,
in terms of $\tilde g_1$, $g_2$ and ${\alpha\over\alpha_c}$. We also plot
this critical surface on the $\tilde g_1-{\alpha\over\alpha_c}$ plane (Fig.~6).
At this approximation level, we find that (i) the critical
line obtained in the continuum theory is modified by the ``lattice'' terms
$\beta_0(r)$ and $\delta_0(r)$ with
$0< r\leq 1$; (ii) $\tilde g_1={1-2\delta_0(r)\over 1+4\beta_0(r)}$ for $\alpha
=\alpha_c$ (the MBLL critical point was $\tilde g_1=1$ for $\alpha =\alpha_c$);
(iii) $\alpha =0$ and $\tilde g_1{\alpha\over\alpha_c} ={4\over 1+8\beta_0(r)}$
(the NJL critical point).
For $\tilde g_1 = 0$, we need to have $2{\alpha \over\alpha_c}\delta_0(r)> 1,
\delta_0(r)> 0$ (see eq.(\ref{critical})), and the critical surface
(\ref{critical})
gives us a critical line at ${\alpha \over\alpha_c}={4\delta_0(r)-1\over
(2\delta_0(r))^2}$. This leads to $2\delta_0(r)> 1$, for which there is no room
for the values of $r$ (see Fig.2).
We also find there is no room
($\tilde g_1 =0$) for the existence of a very small critical value of the gauge
coupling ${\alpha \over\alpha_c}\ll 1$, unless $\delta_0(r)\gg 1$.

\section{Summary}

We begin with a chirally-invariant lattice theory containing gauge interaction
and four-fermion interactions. The NJL self-interaction of only mirror fermions
is added in order to remove mirror fermions from the continuum limit. Tuning
{\it only} one mass parameter (counterterm), we can have the consistent
cancellation between $O({1\over a})$ divergent contributions from mirror
fermions. The SDE for ordinary fermions is free from $O({1\over a})$
divergence.
However, mirror fermions in the lattice region still have finite impacts
$\delta_0(r)\, \beta_0(r)$ and $r(g_2)$ on the SDE for ordinary fermions. The
critical surface for the infrared limit, which is consistent to the critical
line in BLL's model, is thus obtained.

The appearance of composite particles, e.g., Goldstone boson \cite{goldstone},
has been discussed in Ref.\cite{bardeen1} for the continuum theory without
mirror
fermions; in Ref.\cite{xue3} for $\alpha=0$ and in Ref.\cite{xue4} for
$\alpha=0$ and $g_1=0$. In future work, we will present a complete discussion
on this subject.

\newpage  \pagestyle{empty}
\begin{center} \section*{Figure Captions} \end{center}

\vspace*{1cm}

\noindent {\bf Figure 1}: \hspace*{0.5cm}
 The quenched and planar approximated Schwinger-Dyson equation.

\noindent {\bf Figure 2}: \hspace*{0.5cm}
 The function $\beta_0(r)$ in terms of the Wilson parameter $r$.

\noindent {\bf Figure 3}: \hspace*{0.5cm}
 The function $\delta_0(r)$ in terms of the Wilson parameter $r$.

\noindent {\bf Figure 4}: \hspace*{0.5cm}
 The function $r(g_2)$ in terms of $g_2$ ($g_2 > g_2^c\simeq 0.2$).

\noindent {\bf Figure 5}: \hspace*{0.5cm}
 The critical surface (\ref{critical}) in terms of $g_1\,g_2$ and
${\alpha\over\alpha_c}$.

\noindent {\bf Figure 6}: \hspace*{0.5cm}
 The critical surface (\ref{critical}) in  $\tilde g_1-{\alpha\over\alpha_c}$
plane.

}


\begin{thebibliography}{99}

\bibitem{John}
K.~Johnson, M.~Baker and R.~Willey, {\sl Phys.~Rev.} {\bf 136}
(1964) 1111, {\it ibid.} {\bf 163} (1967) 1699;\\
S.L.~Adler and W.A.~Bardeen, {\sl Phys.~Rev.} {\bf D4} (1971) 3045.

\bibitem{maskawa}
T.~Maskawa and H.~Nakajima, {\sl Prog.~Theo.~Phys.} {\bf 52} (1974) 1326 and
{\it ibid.} {\bf 54} (1975) 860.

\bibitem{kugo}
R.~Fukuda and T.~Kugo, {\sl Nucl.~Phys.} {\bf B117} (1976) 250;\\
K.~Higashijima and A.~Nishimura, {\sl Nucl.~Phys.} {\bf B113} (1976) 173;\\
T.~Akiba and T.~Yanagida, {\sl Phys.~Lett.} {\bf B169} (1986) 432.

\bibitem{mira}
V.A.~Miransky and P.I.~Fomin, {\sl Sov.~J.~Part.~Nucl.} {\bf 16} (1985) 203;
P.I.~Fomin, V.~Gusynin, V.A.~Miransky and Yu.~A.~Sitenko,
{\sl Riv.~Nuovo.~Cim.} {\bf 6} (1983)1.

\bibitem{bardeen1}
W.A.~Bardeen, C.T.~Hill and S.~Love,
{\sl Phys. Rev. Lett.} {\bf 56} (1986) 1230,
{\sl Nucl. Phys.} {\bf B273} (1986) 649;\\
W.A.~Bardeen, C.~T.~Hill and M.~Lindner {\sl Phys.\ Rev.} {\bf D41} (1990)
1647.

\bibitem{xue1}
S.-S.~Xue, ``Aspects of dynamical symmetry breaking on a lattice'', MITH 93/20.

\bibitem{nogo}
H.B.~Nielsen and M.~Ninomiya, {\sl Nucl.~Phys.} {\bf B185} (1981) 20, {\it
ibid.} {\bf B193} (1981) 173, {\sl Phys.~Lett.} {\bf B105} (1981) 219.

\bibitem{wilson}
K.~Wilson, in {\it New phenomena in subnuclear physics\/}
(Erice, 1975)
ed.\ A.~Zichichi (Plenum, New York, 1977).

\bibitem{xue}
G.~Preparata and S.-S.~Xue, {\sl Phys.~Lett.} {\bf B264} (1991) 35;
{\sl Nucl.~Phys.} {\bf B26} (Proc.~Suppl.) (1992) 501;
{\sl Nucl.~Phys.} {\bf B30} (Proc.~Suppl.) (1993) 647.

\bibitem{xue3}
G.~Preparata and S.-S.~Xue, ``Emergence of $\bar tt$ condensate and
disappearance of scalar particles'', MITH 93/5, submitted to {\sl Nucl.~Phys.}
{\bf B}.

\bibitem{planck}
C.W.~Misner, K.S.~Thorne and J.A.~Wheeler, {\it Gravitation\/} (Freeman,
San Francisco, 1973).

\bibitem{bardeen2}
W.A.~Bardeen, C.T.~Hill and S.~Love,
{\sl Nucl. Phys.} {\bf B323} (1989) 493,\\
W.A.Bardeen, S.~T.~Love and V.A.~Miransky, {\sl Phys.\ Rev.} {\bf D42} (1990)
3514,\\
A.~Koci\'c, S.~Hands, J.B.~Kogut and E.~Dagotto, {\sl Nucl. Phys.} {\bf B347}
(1990) 217.

\bibitem{wei}
S.~Weinberg, {\sl Phys.~Rev.} {\bf D19} (1979) 1277.

\bibitem{ep}
E.~Eichten and J.~Preskill, {\sl Nucl.~ Phys.} {\bf B268} (1986) 179.

\bibitem{nambu}
Y.~Nambu and G.~Jona-Lasinio, {\sl Phys. Rev.} {\bf 122} (1961) 345.

\bibitem{xue2}
S.-S.~Xue, ``A possible origin of the Wilson lattice fermion'', MITH-93/21 and
hep-lat/9312074.

\bibitem{xue4}
S.-S.~Xue, `` The spontaneous breaking of chiral symmetry
without Goldstone bosons'', MITH 94/5 and hep-lat/9401036.

\bibitem{smit} L. H. Karsten and J. Smit, {\sl Nucl. Phys.} {\bf B144}
(1978) 536.

\bibitem{rome}
A.~Borrelli, L.~Maiani, G.C.~Rossi, R.~Sisto and M. Testa, {\sl Nucl.~ Phys.}
{\bf B333} (1990) 335, {\sl Phys.~Lett.} {\bf B221} (1989) 360.

\bibitem{kondo}
K.-I. Kondo, H. Mino and K. Yamawaki,
Phys. Rev. {\bf D39} (1989) 2430;
K. Yamawaki, {\it in Proc. Johns Hopkins Workshop on
Current Problems in Particle Theory 12, Baltimore, June 8-10, 1988,}
eds. G. Domokos and S. Kovesi-Domokos
(World Scientific Pub. Co., Singapore, 1988);
T. Appelquist, M. Soldate, T. Takeuchi and L.C.R. Wijewardhana,
{\it ibid}.

\bibitem{goldstone}
J.~Goldstone, {\sl Nuovo Cimento} {\bf 19} (1961) 154.

\end{thebibliography}
\end{document}